\documentclass[aps,pra,groupedaddress,showpacs,preprint]{revtex4-1}
\usepackage[english]{babel}
\usepackage{graphicx}

\begin{document}
\title{Potential splitting approach for molecular systems}
\author{E.~Yarevsky$^1$}
\author{S.~L.~Yakovlev$^1$}
\author{N.~Elander$^2$}
\author{{\AA}sa Larson$^2$}

\affiliation{$^1$ St.Petersburg State University, St.Petersburg, 198504 Russian Federation}
\affiliation{$^2$ Stockholm University, 106 91 Stockholm, Sweden}

\begin{abstract}
In order to describe few-body scattering in the case of the Coulomb interaction, an approach based on splitting the reaction potential into a finite range part and a long range tail part is presented.
The solution to the Schr\"odinger equation for the long range tail is used as an incoming wave in an inhomogeneous Schr\"odinger equation with the finite range potential.
The resulting equation with asymptotic outgoing waves is then solved with the exterior complex scaling.
The potential splitting approach is illustrated with calculations of scattering processes in the H${}^+$ -- H${}^+_2$ system considered as the three-body system with one-state electronic potential surface.
\end{abstract}
%

\maketitle

\section{Introduction}
Systems with Coulomb interactions are often found among nuclear, atomic, and molecular systems.
The scattering problem for such systems is of great interest for many physical processes.
The complicated boundary conditions at large distances are a major difficulty for this kind of problems~\cite{FadMerk}.
Several methods have been developed for constructing solutions to the three-body scattering problem (see~\cite{PhysRep.520.135} and references therein).
Some of these avoid using the explicit form of the asymptotic nature of the wave function.

In several recent studies, we have reported a method which is capable to treat correctly the Coulomb scattering problem using exterior complex scaling~\cite{PRA2011, EPL2015, JPB2015, JPhysB2017}.
The key point of this method is splitting the long-range Coulomb potential into the core and tail parts.
The tail part is used to construct the distorted incident wave, which is responsible for the asymptotic Coulomb dynamics.
The core part of the potential generates an inhomogeneous term in the Schr\"odinger equation making possible the application of the exterior complex scaling for solving the equation.
Here we outline the potential splitting approach and present its application to study a molecular system with {\it ab initio} potentials.
Atomic units are used throughout the paper.

\section{Theoretical approach}
%
The three-body quantum systems is described with the Schr\"odinger equation in Jacobi coordinates ${\bf x}_\alpha$, ${\bf y}_\alpha$ with the Hamiltonian
\[
H = T_{{\bf x}_\alpha} + T_{{\bf y}_\alpha} + V({\bf x}_\alpha,{\bf y}_\alpha),
\]
where $T_{{\bf x}_\alpha}$, $T_{{\bf y}_\alpha}$ are the kinetic energy operators, and $V({\bf x}_\alpha,{\bf y}_\alpha)$ is the full interaction in the system. 

Let us consider the scattering of the third particle on the bound state in the pair $\alpha$. 
The reaction potential $V^\alpha({\bf x}_\alpha,{\bf y}_\alpha)$ is defined as
\[
V^\alpha({\bf x}_\alpha,{\bf y}_\alpha) \equiv V({\bf x}_\alpha,{\bf y}_\alpha) - V_{\alpha}({\bf x}_{\alpha})=
\sum_{\beta\neq\alpha} V_{\beta}({\bf x}_{\beta}),
\]
and is split into the sum of the core $V_R$ and the tail $V^R$ parts,
\[
V^{\alpha}({\bf x}_\alpha,{\bf y}_\alpha)   =
V_R({\bf x}_\alpha,{\bf y}_\alpha)   +
V^R({\bf x}_\alpha,{\bf y}_\alpha) ,
\]
where
\begin{eqnarray*}
	\left\{
	\begin{array}{lll}
		V_R~(V^R)&=& V^{\alpha}~ (0), \qquad y_{\alpha} \le R,\\
		V_R~(V^R)&=&  0~(V^{\alpha}), \qquad y_{\alpha} > R.
	\end{array}
	\right.
\end{eqnarray*}

Let us introduce the distorted incident wave $\Psi^R({\bf x}_\alpha,{\bf y}_\alpha)$ as a solution to the scattering problem with the sum of $V_{\alpha}({\bf x}_{\alpha})$ and the tail potential $V^R({\bf x}_\alpha, {\bf y}_\alpha)$:
\begin{equation} \label{Psi-indist}
\left[T_{{\bf x}_\alpha} + T_{{\bf y}_\alpha} + V_{\alpha}({\bf x}_{\alpha}) + V^R({\bf x}_\alpha, {\bf y}_\alpha) - E\right] \Psi^R({\bf x}_\alpha, {\bf y}_\alpha) = 0.
\end{equation}
The total wave function $\Psi$ of the system is represented as a sum 
\[ \Psi =\Phi + \Psi^R , \]
and the function $\Phi$ then satisfies the driven Schr\"{o}dinger equation
\begin{equation} \label{PSA-main}
\left( H - E \right) \Phi = - V_R \Psi^R.
\end{equation}
This is the main equation of the potential splitting approach.
The right hand side of this equation is of finite range with respect to the variable $y_\alpha$.
Thus this equation can be solved numerically with the exterior complex scaling transformation $\quad (W^\theta u)(x) = J(\theta)^{1/2} u[\phi_\theta(x)]$~\cite{ECS, Rescigno1997}.
After application of the ECS to equation~(\ref{PSA-main}), we get 
\begin{equation} \label{PSA-mainCS}
(H(\theta) - E) (W^\theta \Phi) =  - W^\theta V_R \Psi^R,
\end{equation}
where $H(\theta)= W^\theta H (W^\theta)^{-1}$.
As the right hand side of this equation is not analytic for $y_\alpha \leq R$, the exterior complex scaling with the rotation radius $Q \geq R$ has to be applied.
Then $W^\theta \left[V_R \Psi^R\right](r) = 0$  when $y_\alpha$ goes to infinity.

%
In order to solve equation~(\ref{Psi-indist}) and construct $\Psi^R$, let us replace $ V^R({\bf x}_\alpha, {\bf y}_\alpha)$  with its leading term in the incident configuration, $V^R_{\cal C}({y}_\alpha)= \textrm{const}/y_\alpha \,\, (\mbox{or}\,\, 0)$ when $y_\alpha >R\, \, (\mbox{or} \le R)$.
For the potential $V^R_{\cal C}({y}_\alpha)$, the variables approximately separate:
\begin{equation} \label{psi0}
\Psi^R({\bf x}_\alpha, {\bf y}_{\alpha}) \approx  \varphi_{A_0}({\bf x}_{\alpha}) \psi^R_{\cal C}({\bf y}_{\alpha},{\bf p}_{A_0}).
\end{equation}
Here $\varphi_{A_0}({\bf x}_{\alpha})$ is the two-body wave function in the pair $\alpha$ with the set of quantum numbers ${A_0}$, and ${\bf p}_{A_0}$ is the momentum of the incoming particle.
The function $\psi^R_{\cal C}({\bf y}_{\alpha},{\bf p}_{A_0} ) $ in the region ($y_{\alpha} \le R$) is given by
\[
\psi^R_{\cal C}({\bf y}_{\alpha}) =  \frac{4\pi}{p_{A_0}y_{\alpha}}
\sum\limits_{\ell,m}a^R_{\ell}i^\ell {\hat j}_{\ell}(p_{A_0}y_{\alpha})Y_{\ell,m}({\hat p}_{A_0})
Y_{\ell,m}({\hat y}_{\alpha})
\]
with
\[
a^R_\ell = e^{i\sigma_{\ell}}W_R(F_\ell,u^+_\ell)/W_R({\hat j}_\ell, u^+_\ell).
\]
$u^+_{\ell} = e^{-i\sigma_\ell}(G_\ell + i F_\ell)$, $F_\ell(G_\ell)$ are the regular (irregular) Coulomb wave functions,
$W_R(f,g)$ is the Wronskian $f(r)g'(r) - f'(r)g(r)$ calculated at $r = R$.

The full distorted incident wave $\Psi^R$ can then be represented as 
\[ \Psi^R = \Psi^R_0 + \Psi^R_1 , \]
where $\Psi^R_0$ is the distorted wave~(\ref{psi0}), and $\Psi^R_1$ satisfies the inhomogeneous equation:
\[
\left(T_{{\bf x}_\alpha} + T_{{\bf y}_\alpha} - V_\alpha({\bf x}_\alpha) + V^R - E\right)\Psi^R_1 = -(V^R-V^R_{\cal C}) \Psi^R_0 .
\]
In the region where $\varphi_{A_0}({\bf x}_{\alpha})$ is not negligible, one obtains 
\[
V^R({\bf x}_\alpha, {\bf y}_\alpha)-V^R_{\cal C}(y_{\alpha})\sim O\left(y^{-2}_{\alpha}\right).
\]
The non-Coulomb tail of this remainder potential can be truncated at some $R' > R$~\cite{Rescigno1997}.
The error because of this truncation goes to zero with increase in $R'$.

The solution $\Phi$ of the problem is then represented as
\[ \Phi = \Phi_0 + \Phi_1, \]
where the functions $\Phi_0$, $\Phi_1$ are the solutions to the equations
\[
\left(H - E\right) \Phi_k = - V_R \Psi^R_k, \quad k=0,1.
\]
The total wave function $\Psi$ is the sum
\[
\Psi = \Psi^R_0 + \Phi_0 + \Psi^R_1 + \Phi_1 ,
\]
where the two last terms vanish when $R \to \infty$.
For moderate values of $R$, however, their contributions might not be negligible.



In order to find scattering amplitudes and cross sections with the wave function, the asymptotic form of the scattered wave function at large distances is used~\cite{FadMerk}:
\begin{equation} \label{originBC}
\Phi \sim
\sum\limits_{n,\ell}\frac{R_{n,\ell}(x_\alpha)}{x_\alpha}Q_n(y_\alpha, p_n)
Y_{\ell,0}(\theta_\alpha,0)  {\tilde{A}}_{n,\ell}
+ B(x_\alpha, y_\alpha, \theta_\alpha) ,
\end{equation}
where the function $B(x_\alpha, y_\alpha, \theta_\alpha)$ represents the three-body ionization term.
For large hyperradius $\rho_\alpha=\sqrt{x_\alpha^2 + y_\alpha^2}$, it decreases as $B(x_\alpha, y_\alpha, \theta_\alpha) \sim \rho_\alpha^{-1/2}$.
The total state-to-state $(n_i,\ell_i) \to (n,\ell)$ scattering amplitude ${A}_{n,\ell}$ is split into three terms
\begin{equation} \label{totalAmpl}
{A}_{n,\ell} = [A_0^R]_{ n, \ell}+[A_1^R]_{n,\ell}+{\tilde{A}}_{n,\ell}.
\end{equation}
The term $[A_0^R]_{ n, \ell}$ corresponds to the function $\Psi^R_0$ and is calculated explicitly.
The terms $[A_1^R]_{n,\ell}$ and ${\tilde{A}}_{n,\ell}$ correspond to the functions $\Psi^R_1$ and $\Phi=\Phi_0 + \Phi_1$, respectively.
Projecting the representation~(\ref{originBC}) on the two body wave functions, the local representation for the partial amplitudes ${\tilde {\cal A}}_{n,\ell}$ can be derived~\cite{JPhysB2017}:
\begin{equation}
{\tilde {A}}_{n,\ell} \approx \frac{Q_n^{-1}(y_{\alpha},p_n)}{2\pi}
\int\limits_0^{\infty}dx_\alpha \int\limits_0^{\pi} \sin\theta_\alpha d\theta_\alpha  
x_{\alpha} R_{n,\ell}(x_\alpha) 
\Phi (x_\alpha, y_{\alpha}, \theta_\alpha)Y_{\ell, 0}(\theta_\alpha,0)  .
\end{equation}

To summarize, the solution of the scattering problem becomes a two-step procedure.
At the first step, the driven equation with the exterior complex rotation~(\ref{PSA-mainCS}) is solved.
The zero boundary conditions at infinity are used to construct the solution.
At the second step, the scattering amplitudes are calculated with the representation~(\ref{totalAmpl}).
This is done inside the non-rotated region, so the original boundary conditions~(\ref{originBC}) are used.

%
\section{Application of the potential splitting method}

Molecular systems cannot be studied with few-body methods as the total number of particles is too large.
To apply such methods, additional approximations are necessary. 
The most obvious one is the Born-Oppenheimer approximation where the electron degrees of freedom do not participate in the dynamical equations but are averaged to the potential energy surfaces.
For accurate calculation of processes, {\it ab-initial} potentials calculated with quantum chemistry approaches should be used.
In the case of three-body systems, these potentials depend on all three interparticle distances, and are given numerically.
The exterior complex scaling approach can be used to calculate scattering processes with this type of potentials provided that the rotation radius is larger than the interparticle distance where the potential is numerically calculated.
For systems with asymptotic Coulomb interaction, the potential splitting approach should be additionally used. 

In this work, we have considered the H${}^+$ -- H${}_2^+$ scattering.
The H${}_2^+$ ion is carefully studied~\cite{H2plus}.
Its ground state energy is -0.59711~a.u., and there exist 20 bound states for total zero angular momentum, known with the very high accuracy~\cite{H2plus}.

The potential energy surface $U(r_1,r_2,\theta)$ of electronic ground state of H${}_3^{2+}$ depends on the two bond-lengths $r_1$, $r_2$, and the angle $\theta$ between them.
It is computed using the aug-cc-pVQZ basis set of Dunning~\cite{Dunning1989}. 
Using the Full Configuration Interaction method, the three lowest electronic states in ${}^2$A’ sym\-met\-ry are computed in order to verify that the electronic ground state is well separated in energy from the excited electronic states and that the non-adiabatic effects can be neglected. 
This is the case for the region of the potential energy surface probed in the H${}^+$ -- H${}^+_2$ collisions studied here at relative low collision energies.
The {\it ab initio} calculations are carried out using internal coordinates where the bond-lengths are varied in the range $0.8 a_0 \leq r_i \leq 20.0 a_0$ and the angle between the two bond-lengths is varied in $[0, \pi]$.
The potential energy surface is computed on an product grid, where 33 values of the internuclear bond-lengths and 37 values for the angle are used. 
The {\it ab initio} calculations are carried out using the molpro program~\cite{Molpro}.
For the regions where two nuclei come close together and the asymptotic regions at large internuclear distances, the {\it ab initio} potential energy surface is extrapolated. 

In order to make the extrapolation, let us introduce the function
\begin{equation} \label{regulariz}
V(r_1,r_2,\theta) = U(r_1,r_2,\theta) - \sum_{i=1}^3 E_{H_2^+}(r_i).
\end{equation}
where $r_3$ is calculated as $r_3=\sqrt{r_1^2+r_2^2- 2 r_1 r_2 \cos\theta}$.
The energy $E_{H_2^+}(r)$ is the energy of the Coulomb two-centre problem with the electron and two charges +1 each placed at the distance $r$.
It can be calculated both with the quantum chemistry approach (the same as is used for the H${}_3^{2+}$ calculations) and with the semi-exact approach~\cite{odkil}.
The functions $E_{H_2^+}(r_i)$ are more local compare to $1/r_i$, i.e. they decrease much faster for large distances.
The function $V(r_1,r_2,\theta)$ has no singularities at $r_i=0$ and hence is much easier to interpolate and extrapolate.

Firstly, the points with $r_i=0$, $i=1,2,3$, are added to the grid for $V(r_1,r_2,\theta)$.
Namely, let a chosen distance $r_i$ be very small, $r_i=\varepsilon$.
The energy $E_{H_2^+}(\varepsilon)$ is then expanded in powers of $\varepsilon$ as
\[
E_{H_2^+}(\varepsilon) = 1/\varepsilon - 2 + O(\varepsilon^2).
\]
On the other hand, as $\varepsilon \ll 1$, the whole H$_3^{++}$ system can now be considered as an electron in the field of two Coulomb centers with charges +2 and +1 placed at a distance $y$.
The total energy $E_{p^+(2p)^{++}e}$ of the H$_3^{++}$ system in this configuration is equal to 
\[
E_{p^+(2p)^{++}e}(y) = W_{+1+2e}(y) + 1/\varepsilon,
\]
where $W_{+1+2e}(y)$ is the energy of the two Coulomb center problem which can be calculated with the ODKIL program modified according to formulas given in paper~\cite{odkil}.
Then
\[
E_{p^+(2p)^{++}e}(y) = W_{+1+2e}(y) + 1/\varepsilon - 2 + 2 = W_{+1+2e}(y) + 2 + E_{H_2^+}(\varepsilon).
\]
Taking into account Eq.~(\ref{regulariz}), we find for the regularized function $V(r_i,r_j,r_k)$ at $r_i=0$ the following value
\begin{equation} \label{reg_1zero}
V = W_{+1+2e}(r_j) + 2 - E_{H_2^+}(r_j) - E_{H_2^+}(r_k),
\end{equation}
where $r_j=r_k$, and $i \neq j \neq k$.

When all three distances $r_i$ approach zero, the whole system approaches a hydrogen-like atom with the +3 charge in the united atom approximation.
Its energy $E_0$ is written as
\[
E_0 = \sum_{i=1}^3 {1 \over r_i} - \frac{9}{2} +O(\max_i{(r_i)}).
\]
For the regularized function $V(r_i,r_j,r_k)$ at zero distances, this gives
\[
V(0,0,\theta) = \frac{3}{2}.
\]
The latter value coincides with the value calculated from Eq.~(\ref{reg_1zero}) in the limit $r_j=r_k \to 0$.

Using Eq.~(\ref{reg_1zero}), missing points for short distances $r_1$, $r_2$ can be filled in so the bond lengths span the interval $0 \leq r_i \leq 20.0 a_0$.
Now if $\max_i{(r_i)} \leq 20.0 a_0$, the potential energy surface can be calculated at an arbitrary bond lengths with the 3D spline interpolation on the given grid.

\paragraph{Calculations for the asymptotic region}
If one or few distances are outside the numerical grid, $r_i \geq RB$, the interpolation procedure may not be used so an extrapolation has to be devised. 
It is done in the following way:
\begin{itemize}
	\item[1.] Sort the distances in the ascending order, so that $\overline{r}_1 \leq \overline{r}_2 \leq \overline{r}_3$.
	\item[2.] If $\overline{r}_1 \leq \overline{r}_2 \leq RB \leq \overline{r}_3$, the PES calculated with the 3D spline interpolation from the given numerical grid.
	Although the value $\overline{r}_3 \geq RB$, the energy for this configuration is present in the numerical grid.
	\item[3.] All three particles are far away from each other, $RB \leq \overline{r}_1$. 
	For this arrangement, the minimum of energy is found in the configuration with the electron located near the proton 3, as this gives the lowest repulsive Coulomb energy while the attractive polarization energies are relatively small due to the large distances.
	The electric field at the particle 3 position is given by the vector sum of the fields of other particles:
	\[
	|\vec{E}|^2 = \frac{1}{\overline{r}_1^4} + \frac{1}{\overline{r}_2^4} + \frac{2 \cos\theta_{12}} {\overline{r}_1^2 \overline{r}_2^2},
	\]
	where 
	$\cos\theta_{12} =(\overline{r}_1^2+ \overline{r}_2^2 - \overline{r}_3^2)/(2 \overline{r}_1 \overline{r}_2)$.
	Then the interaction of the induced dipole with the fields plus the Coulomb energy gives the interaction energy
	\begin{equation}
	U(\overline{r}_1, \overline{r}_2, \overline{r}_3) = 
	\frac{1}{\overline{r}_3} +
	\left(-\frac{\alpha}{\overline{r}_1^2} -\frac{\alpha}{\overline{r}_2^2} \right)
	\sqrt{\frac{1}{\overline{r}_1^4} + \frac{1}{\overline{r}_2^4} + \frac{2 \cos\theta_{12}} {\overline{r}_1^2 \overline{r}_2^2} }.
	\end{equation}
	Here $\alpha=2.5344$ stands for the coefficient, fitted from the 
	$E_{H_2^+}(r)$ energy at large $r$.
	
	\item[4.] The configuration $\overline{r}_1 \leq RB \leq \overline{r}_2 \leq \overline{r}_3$ corresponds to the situation when the particle 1 is far away from the pair of 2 and 3.
	The main interaction here is the Coulomb interaction of the particle 1 and the pair, perturbed with the $\beta/\overline{r}_3^2$ term as the position of the charge distribution center is unknown.
	Assuming this asymptotic behavior, its coefficient is defined from the numerical data at the largest distance available.
	When particle 1 approaches particle 2 along the line in the $\vec{\overline{r}}_3$ direction, this distance can be $\overline{r}_2 = RB$ if the corresponding value $\hat{r}_3 > RB$, where $\hat{r}_3= \overline{r}_1 \cos\theta_{13} + \sqrt{\overline{r}_1^2 \cos^2\theta_{13} + RB^2 - \overline{r}_1^2}$, 
	$\cos\theta_{13} =(\overline{r}_1^2+ \overline{r}_3^2 - \overline{r}_2^2)/(2 \overline{r}_1 \overline{r}_3)$, 
	or
	$\overline{r}_3 = RB$, if not.
	Hence, the total potential is represented in the form
	\begin{equation} \label{pot_asym_pair}
	U(\overline{r}_1, \overline{r}_2, \overline{r}_3) = 
	E_{H_2^+}(\overline{r}_1) + \frac{1}{\overline{r}_3} +
	\frac{\beta}{\overline{r}_3^2} .
	\end{equation}
	The parameter $\beta$ is defined from the relation
	\[
	U(\overline{r}_1, RB, \hat{r}_3) = 
	E_{H_2^+}(\overline{r}_1) + \frac{1}{\hat{r}_3} +
	\frac{\beta}{\hat{r}_3^2}
	\]
	for $\hat{r}_3 > RB$, and from the relation
	\[
	U(\overline{r}_1, \hat{r}_2, RB) = 
	E_{H_2^+}(\overline{r}_1) + \frac{1}{RB} +
	\frac{\beta}{RB^2}
	\]
	otherwise.
	Here $\hat{r}_2= \sqrt{\overline{r}_1^2 + RB^2 - 2 \overline{r}_1 RB \cos\theta_{13}} > RB$.
	The values $U(\overline{r}_1, RB, \hat{r}_3)$ and $U(\overline{r}_1, \hat{r}_2, RB)$ can be determined from the numerical grid as two distances $\overline{r}_1$ and $RB$ are not greater than $RB$.
	The potential (\ref{pot_asym_pair}) depends on $\overline{r}_2$ implicitly as this value is used for the calculation of $\beta$ parameter.
\end{itemize}

We have calculated the elastic and excitation scattering cross sections H$_2^+$($v=0,J=0$)+H$^+$ $\to$ H$_2^+$($v',J'=0$)+H$^+$ with the constructed potential energy surface.
The numerical solution of the driven Schr\"odinger equation~(\ref{PSA-mainCS}) is performed by the finite element method (FEM), which is described in details in~\cite{ELY-helium}.
In the calculations, we use a rectangular product grid.
For the reaction coordinate $y_\alpha$, five finite elements have been used at short distance [0–4]~a.u., 44 elements for intermediate region, and ten elements of total length 40~a.u. for the discretization beyond the splitting point $R=$~31~a.u.
For the coordinate $x_\alpha$, 19, 9, and 4 elements respectively have been used for the regions mentioned above.
One element has been used for the angular variable~$\theta_\alpha$.

\begin{figure}[t]
	\centering \includegraphics[width=0.75\linewidth]{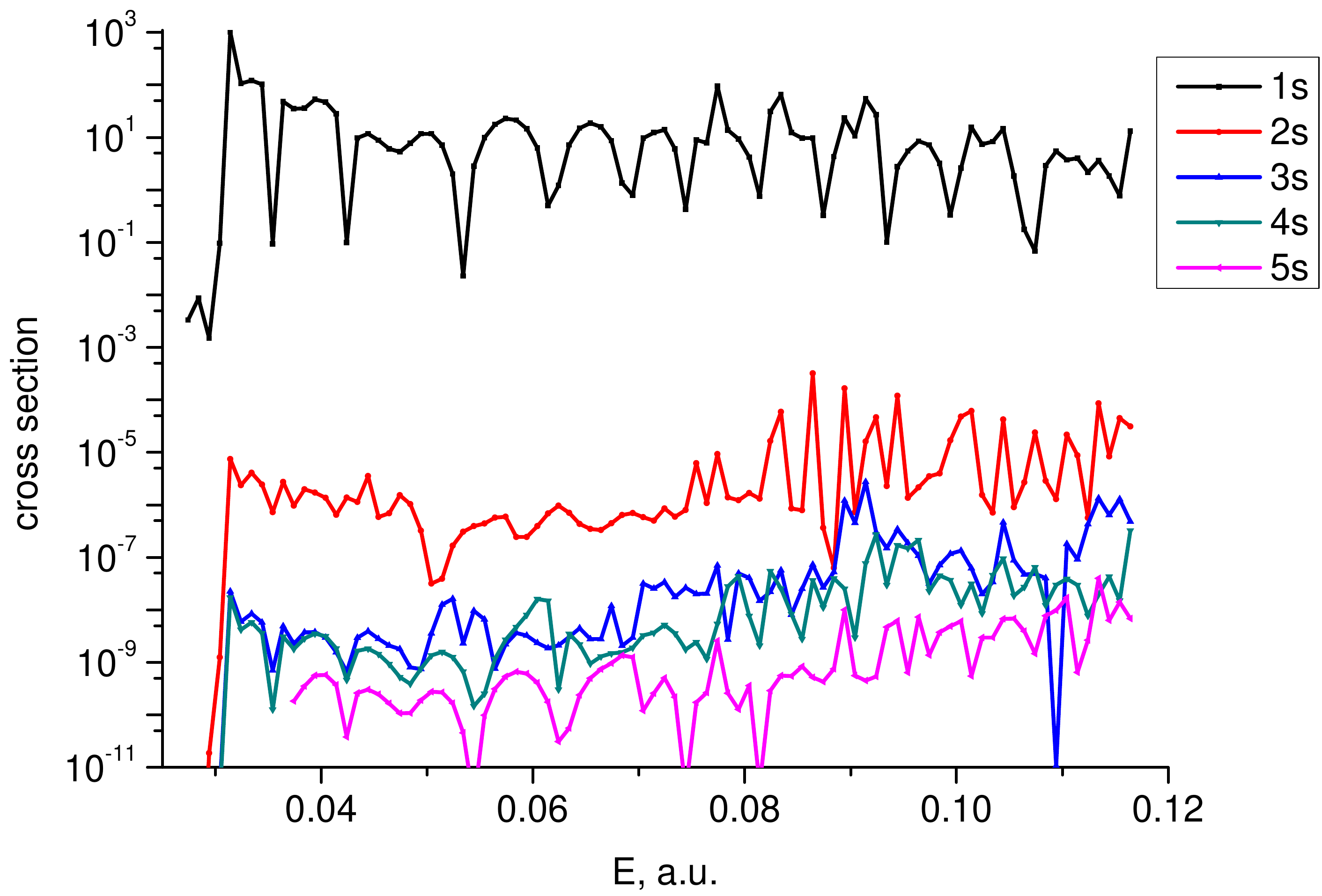}
	\caption{The elastic and excitation cross sections for the H${}^+$ -- H${}_2^+$ scattering.}
	\label{fig-molec}
\end{figure}

Our results for the elastic and excitation H$_2^+$($v=0,J=0$) $\to$ H$_2^+$($v',J'=0$) cross sections for the H${}^+$ -- H${}_2^+$ scattering are presented in figure~\ref{fig-molec}, where the energy $E$ is the incident energy of H${}^+$.
The structure in the cross section appears because of large number of states in the H${}_2^+$ molecule.

\section{Conclusions.}
We have proposed the mathematically sound approach for calculations of scattering processes.
The potential splitting approach allows for the solution of the scattering problem with the Coulomb interaction. 
Besides systems with explicitly given analytical interactions, molecular systems with numerically defined {\it ab initio} potentials can be studied with the combined exterior complex scaling and splitting potential approaches.

\section{Acknowledgements.} 
Financial support from the RFBR grant No. 18-02-00492 is acknowledged.
{\AA}L acknowledges support from the Swedish Research Council under project number 2014-4164.
The calculations were carried out using the facilities of the ``Computational Center of SPbSU''.

\end{document}